# NFT Appraisal Prediction: Utilizing Search Trends, Public Market Data, Linear Regression and Recurrent Neural Networks

**Shrey Jain, Camille Bruckmann, Chase McDougall**

Faculty of Engineering

University of Toronto

{shreyd.jain,camille.bruckmann, chase.mcgdougall}@mail.utoronto.ca

## Abstract

Non-Fungible Tokens (NFTs) are transforming human coordination across economic, governance, and social protocols of Web3 [1] [2]. Despite their nascency, NFTs have a market cap of $50B and are becoming a key element of ownership in the digital world [3].

Similar to traditional equity markets, NFT markets are volatile and challenging to speculate upon. Forecasting future values of NFTs are important for individual financial security and investment making as they become a primitive asset class for investors. The NFT market is highly heterogeneous with 1% of assets being traded for over $1500 and 75% being traded for less than $15 [4]. Additionally, only 32,000 wallets hold 80% of the market cap of the NFT market making the inequities clear and ability for rapid changes to markets more easily accessible [5].

In this paper we investigate the correlation between NFT valuations and various features from three primary categories: **public market data**, **NFT metadata**, and **social trends data**. These data sources were chosen such that we could draw connections to more traditional investment classes as well as make an effort to quantify qualitative data, such as sentiment and engagement through social media and search trends.

Additionally, this data was passed through various machine learning models, each with varying goals. A linear regression model was used to identify the features most strongly correlated with an NFT's valuation as well as to identify redundancy among features; furthermore, various iterations of RNNs were investigated including: LSTMs, GRUs, and Multivariate versions, in an effort to project future NFT valuations.

# Prior Work

The nascency of the NFT space limits the quantity of prior work that has been done in this space. Additionally, the venues for peer-reviewed journals in the social elements of the blockchain (not cryptography), are yet to exist. As such, the works we reference will be gathered from the most well-reviewed sources, but will also stem from credible authors' blogs and company posts.

## Mapping the NFT revolution: market trends, trade networks, and visual features [13]

This paper aims to demystify the overall structure of the Non Fungible Token (NFT) market and provides a framework for quantifying its evolution. The authors analyzed data concerning 6.1 million trades and 4.7 million NFTs between 2017 and 2021, obtained primarily from Ethereum and WAX blockchains. The results of this paper include the statistical properties of the market, a network of interactions between traders (linked by buyer and seller), and a clustering of objects by visual features and collections. The paper also proposes a linear regression model with features based on these results to predict NFT prices.

From the top five collections in the NFT market, the most exchanged NFTs belong to the categories Games, Collectibles, and Art, which account for 44%, 38%, and 10% respectively of transactions. In terms of market volume, the Art category has dominated, contributing ~71% of the total transaction volume. With regards to the relationship between traders, the top 10% of traders (measured by their number of purchases and sales) perform 85% of all transactions. Furthermore, traders specialized in a collection tend to buy and sell NFTs with other traders specialized in the same collection. AlexNet, a convolutional neural network, was used to produce dense vector representations of images, and principal component analysis (PCA) was used to study the similarity between 1.25 million different NFTs. It was found that the first five principal components account for 38.3% of the total variance and are used to test the capacity of visual features for predicting NFT sales.

A linear regression model with a least squares loss was developed to predict the price of primary and secondary sales, calculated considering the data preceding the day of the NFT's primary sale. Eleven features, partitioned into three groups, were used to characterize the NFTs. They are summarized in Table 3.

| Group 1: Network centrality (4) | Group 2: Visual features (5) | Group 3: Sale history (2) |
|---|---|---|
| - Degree centrality of seller<br>- PageRank centrality of seller<br>- Degree centrality of buyer<br>- PageRange centrality of buyer | - Five PCA components extracted from the AlexNet vector of the NFT. | - Median price of primary and secondary sales made in the collection of interest.<br>- The prior probability of secondary sale. |

**Table 1:** Summary of features used to train a linear regression model to predict price of primary and secondary sales of an NFT.

It was determined that the median sale price of NFTs in the collection predicts more than half of the variance of price of future primary and secondary sales, and that the prediction is more accurate when the median of the past sale price is calculated over a recent time window preceding the primary sale. Moreover, centrality measures of the buyer and seller in the trader network as well as visual features of the object linked to the NFT explain roughly one-fifth of the variance when used in combination. Several limitations of this study include data gathered from NFT marketplaces instead of directly from Ethereum or WAX blockchains, causing independent NFT producers to be left unaccounted for. Other influences on market behavior, such as social media, were not included in the analysis.

## The NFT Hype: What Draws Attention to Non-Fungible Tokens? [15]

This paper focuses on utilizing vector autoregressive models (VARs) to show that core cryptocurrencies, namely Bitcoin (BTC) and Ether (ETH) draw the most attention towards predicting future NFT price.

This team utilizes the S&P 500, google search trends, and the prices of cryptocurrencies as indicators for future price of an NFT. This team highlights that google search trend data is associated with major cryptocurrency returns and NFT collections. In addition to VARs this team uses wavelet coherence techniques to investigate co-movement between cryptocurrency returns and NFT levels of attention. The results of this paper show that there is no significant relationship between Ether returns and attention to NFTs but there is a relationship between Bitcoin and the prediction of an NFT.

## TweetBoost: Influence of Social Media on NFT Valuation [16]

This paper aims to answer two main questions: a) What is the relationship between user activity on Twitter and price on OpenSea? b) Can we predict NFT value using signals obtained from Twitter and OpenSea, and identify which features have the greatest impact on the prediction?

While answering this question, the paper seeks to create one of the first NFT datasets consisting of both OpenSea and Twitter data. Using both a Binary and Multi-classification model to first predict whether or not the NFT will be profitable and then classifying the profitable NFTs into varying price brackets. Kapoor et al. concluded that in adding Twitter data to their feature set they were able to increase the accuracy of their model by 6% when compared to a model only using data from NFT platforms (such as OpenSea). This paper gives insights into additional training strategies and features for use within our predictive model.

# Data Collection and Processing

## NFT Data

Today, there are more NFTs on OpenSea and other NFT marketplaces than there were websites in 2010 [30]. As such, querying and searching data related to NFTs has become a challenging problem due to the volume and variation in data types.

While it may seem trivial to index historical data on the Ethereum blockchain related to digital assets, we found this to be a challenge with existing APIs as compared to traditional finance market APIs (i.e. Yahoo finance). Ethereum nodes today do not natively store transactions by wallet address, and this must be done with an indexer. Currently, the indexers that have been built to query the Ethereum blockchain have high developer friction, are fragmented, or are prohibitively expensive for this project.

In order to get the data we were interested in gathering related to NFTs we needed to explore more APIs than we initially intended to in our project proposal. The APIs we explored include [1]:

- OpenSea NFT API
- Covalent NFT historical data API
- Etherscan API
- Coingecko API
- Moralis API

We go through a more detailed process of how we converged to using the Covalent API in the Appendix. The Covalent API has become one of the more reputable teams working on indexing blockchain data across multiple chains (i.e., Ethereum, Binance, Polygon, Solana, Ronin, etc.). Covalent is backed by some of the top crypto venture capital firms in the world and have built a strong reputation within the crypto developer community which made our team feel more confident in the quality of data used in this project.

As seen in the `NFTValuation.ipynb` file, we use the Covalent API to get the following data: date; average price of the NFT on a given day; and number of NFTs sold on a given day.

The set of NFT collections that we pulled from, as highlighted in our project proposal include: Geisha Tea House; BAYC; Cryptopunks; Doodles; Azuki; Deadfellaz; Gutter Cat Gang; Sup Ducks; Cyber Kongs; Creature World; Cool Cats; World of Women; Alien Frens; Lazy Lions. [2]

---

[1] To get more specific details about the data dictionaries that each of these APIs yield, refer to the Appendix. We also highlight the specific challenges with each of these APIs in the Appendix.

[2] See Appendix for a list of the contract addresses that we pulled from Etherscan to feed to the Covalent API.

## Public Market Data

To gather public market data, we used the commonly referenced Yahoo Finance API. The Yahoo Finance API is a widely used tool for financial data that has led to it becoming extremely frictionless for developers. As noted in our project proposal, we want to concatenate both NFT specific data with public market data to predict the future price of an NFT.

By adding all of the tickers (an abbreviation used to uniquely identify publicly traded shares of a particular stock on a particular stock market), we are able to put all of the public market data into a dataframe that can be easily concatenated with our NFT data. This is all clearly documented in our `NFTValuation.ipynb` file.

## Google Trends Data

With the majority of an NFT's value being tied to *Hype* surrounding a particular project, it is important for us to determine a way to quantify this metric. One such method is in using Google Trends data to track the relative search volume of a collections name over time. The PyTrends API allows us to extract this information for concatenation with our dataset.

## Significant Event Data

Other hard to quantify variables include news and announcements related to the collections we are forecasting. For example when the Bored Ape Yacht Club announces a new spin-off or a scammer targets the collection we can see a positive or negative impact on pricing, respectively. We aim to quantify this through the inclusion of an Event feature with positive values indicating a positive price correlation and negative values indicating a negative price correlation.

## Aggregated Data Dictionary

With the NFT, public market data, and Google Trends data, we were able to build the following data dictionary that would prepare the data to be fed into a neural network.

| Variable | Description | |
|---|---|---|
| `opening_date` | Date of which information is being pulled. | |
| `average_volume_quote_day` | Average price of the NFT as of `opening_date`. | |
| `unique_token_ids_sold_count` | The number of NFTs from a given collection sold in one day. | |
| `ETH_USD` | Closing price of ETH token. | **X (input data)** |
| `BTC_USD` | Closing price of Bitcoin. | |
| `GC=F` | Closing price of gold. | |
| `^GSPC` | Closing S&P value. | |
| `^DJI` | Closing Dow Jones value. | |
| `^NDX` | Closing Nasdaq 100 value. | |
| `MSFT` | Closing Microsoft stock price. | |
| `AAPL` | Closing Apple stock price. | |
| `NFLX` | Closing Netflix stock price. | |
| `TSLA` | Closing Tesla stock price. | |
| `AMZN` | Closing Amazon stock price. | |
| `FB` | Closing Meta stock price. | |
| `Relative Search Volume` | Relative google search volume for collection name on a scale from 0-100 | |
| `Events` | -1,0,1 indicating bad news, no news, and good news respectively | |
| `Gas` | A measure of network traffic, which indicates the transaction fee of purchase | |
| `average_volume_quote _day[opening_date+1]` | Average price of the NFT as of `opening_date + 1` (the next day's price). | **Y (labels)** |

**Table 2.** This table is our data dictionary that contains all of the relevant information needed to feed into our linear or recurrent models.

## Data Adaptations

Due to fundamental differences between the crypto market and traditional stock market, additional adjustments were required in order to have data that is properly aligned. In particular, the NFT and crypto markets are open 24 hours a day, 7 days a week, while traditional markets are only open Monday to Friday during trading hours. In order to adapt our NFT data to be usable in conjunction with stock information, we decided to only use the NFT data associated with days that have corresponding stock data. The same adjustment was required for information that has data for every day of the year, such as Google Trends data.

# Limitations / Roadblocks

## Twitter API

We intended to incorporate Twitter activity related to a collection in order to aid in measuring *Hype* of that collection; however, while we were able to obtain access to the Twitter API, we were unable to gain access to the endpoint(s) necessary for collecting the volume data we were interested in. Specifically, we require access to the twitter `/search/all` endpoint to query specific tweet information, which requires elevated permissions. Through access to this endpoint we planned to gather data relating to the volume of collection related tweets as a feature, which we have proxied through the use of Google Trends data. Alternative uses for this access would be to perform sentiment analysis on the tweets surrounding the collection; however, this is something we were unable to proxy.

## OpenSea Events API

Exploring "owner similarity", the similarity of owners between NFT collections, was a metric our team wanted to explore further using the OpenSea Events API. As noted in Nadini et al.'s, there is a small network of successful NFT traders that exhibit similar bimodal behavior (either trading low volume of high-priced assets or high volume of low-priced assets). For each timestep to our model, we wanted to build a list of K successful collections (as determined by market cap) and compute a similarity score for a collection N (that is not in the set K) of the NFT owners. We would then be able to understand the relationship between the wallet owners and the predictive power of the NFT collection. This was largely influenced by the fact that 80% of the market cap of NFTs is held within 10% of the wallets (for Ethereum based NFTs).

# Baseline Model Implementations

## Recurrent Neural Network

Our first approach at predicting the average price of a BAYC NFT involved building a recurrent neural network (RNN). RNNs are deep learning models that take in sequential data as input which makes them well suited for time series predictions. The data fed into our model is shown in Figure 1. Before converging on the final model, several different types of RNNs were built to predict the average price of a BAYC NFT. These include vanilla RNNs, LSTMs, GRUs, etc.

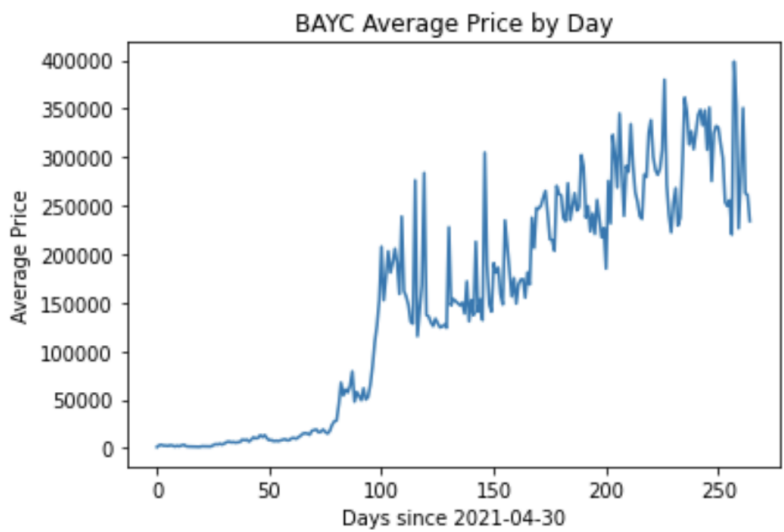

**Figure 1:** A plot of the time series data fed into the RNN.

The RNN with GRU cells had the best validation accuracy out of the other types of RNNs tested. A grid search was conducted to determine the best hyperparameters which are summarized in the table below:

| Number of GRU Blocks | Hidden Dimension Size | Learning Rate | Sequence Length |
|---|---|---|---|
| 3 | 16 | 0.01 | 5 |

**Table 3.** RNN hyperparameters.

The RNN was fed the average price of a BAYC NFT over time. After training and testing, the RNN had an average percent error of 80% on the training set, 168% on the validation set, and 185% on the test set. Figure 2 shows the RNN's training prediction as well as the true price of the NFT over time.

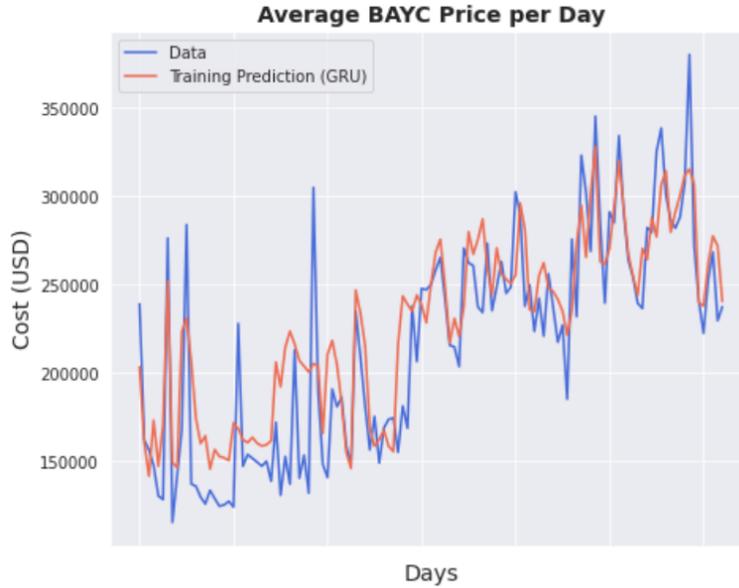

**Figure 2:** A plot of the average BAYC price per day. The blue line indicates the true value, while the red indicates our training predictions.

These results are not unexpected considering the volatility of NFT prices, however, we were unsatisfied with this result. Upon completing our thorough investigation of RNNs for price prediction, we decided to pivot away from predicting the absolute price of an NFT and to instead focus on discovering the underlying factors that influence NFT prices.

## Linear Regression

Inspired by the linear models found in the works of Nadini et. al. (Mapping the NFT Revolution) and Kapoor et. al. (TweetBoost), a linear regression model was used to determine which predictors influence the price of an NFT. In our first model, we fed the 16 different predictors listed in Table 2. We assessed the appropriateness of the predictors by evaluating different model diagnostics. These included looking at the standard error of the regression coefficients, the associated t statistics, and the Variance-Inflation Factors (VIF) associated with predictor multicollinearity. These results for the initial model can be seen below in Table 4.

| Predictor | Coeff | VIF | t | P > |t| | Confidence Interval |
|---|---|---|---|---|---|
| Days since release | 958.6 | 548.9 | 1.166 | 0.246 | -670.4, 2587.6 |
| Avg. NFT price | 0.09 | 41.8 | 1.053 | 0.294 | -0.082, 0.267 |
| # of NFTs sold | 987.6 | 4.9 | 4.163 | 0.000 | 517.6, 1457.5 |
| Gas | 11.5 | 286.0 | 0.732 | 0.466 | -19.7, 42.8 |
| ETH-USD | -14.2 | 749.6 | -0.541 | 0.590 | -66.0, 37.7 |
| BTC-USD | -0.6 | 705.0 | -0.352 | 0.726 | -4.2, 3.0 |
| Gold | 165.8 | 3309.4 | 1.335 | 0.185 | -80.3, 411.9 |
| S&P | 72.6 | 531499.6 | 0.136 | 0.892 | -987.9, 1133.2 |
| Dow Jones | -27.7 | 140946.3 | -0.743 | 0.459 | -101.7, 46.3 |
| NASDAQ | -8.0 | 142726.0 | -0.097 | 0.923 | -170.9, 154.9 |
| MSFT | 811.9 | 8842.1 | 0.792 | 0.430 | -1218.0, 2841.9 |
| AAPL | -553.0 | 2677.4 | -0.499 | 0.618 | -2746.4, 1640.4 |
| NFLX | -57.5 | 797.6 | -0.336 | 0.738 | -396.4, 281.5 |
| TSLA | 30.0 | 456.9 | 0.368 | 0.713 | -131.1, 191.2 |
| AMZN | 28.9 | 2311.5 | 0.598 | 0.551 | -66.8, 124.6 |
| FB | 238.9 | 838.3 | 0.803 | 0.424 | -350.4, 828.3 |

**Table 4**: The coefficient, VIF, t-statistic, p value, and confidence interval for each predictor.

As can be seen above in Table 4, multiple predictors exhibited significant multicollinearity, indicating high correlations between the predictors. This will make the regression surface unstable and artificially inflate the variances of the predictors, making all diagnostics tests and confidence intervals unreliable. This was expected as several predictors are highly correlated, such as the NASDAQ, S&P, and Dow Jones prices, and the individual stocks.

It can also be seen in the table that several predictors are not significant to the 0.05 level, likely caused by multicollinearity. It appears that only a few predictors accounted for the majority of the model quality, indicating that several can be cut. The initial model has an adjusted $R^2$ value of 0.7, indicating overall high model quality.

To improve model quality and reliability, two additional predictors were added, Events and Relative Volume Search, to try to capture the effect of "hype" on the price of the NFT. After adding these predictors and recalculating the VIF factors, we started to remove predictors that exhibited significant multicollinearity. The predictors with high p values were kept and the VIF factors were reassessed. The final model contains four predictors, all with relatively low VIFs that are considered acceptable in data analysis. Out of these predictors, two were significant to a 0.05 level, and the other two to a 0.1 level. The price the previous day, and the daily volume were by far the most influential predictors on the NFT price the following day. This final model had an adjusted $R^2$ value of 0.69, indicating virtually all model quality was retained from the initial model, but redundant predictors were removed. This final model and the predictor information can be seen below in Table 5.

| Predictor | Coeff | VIF | t | P > \|t\| | Confidence Interval |
|---|---|---|---|---|---|
| Average NFT Price | 1168.0 | 6.6 | 11.8 | 0.000 | 972.5, 1363.5 |
| Number of NFTs sold | 903.3 | 2.7 | 4.7 | 0.000 | 525.2, 1281.4 |
| Gas | 11.0 | 5.3 | 1.9 | 0.056 | -0.3, 22.2 |
| Relative Search Volume | 407.3 | 4.3 | 1.8 | 0.072 | -37.5, 853.4 |

**Table 5:** The coefficient, VIF, t-statistic, p value, and confidence interval for the final set of predictors.

The model was also evaluated on a test set, to verify the model did not overfit, and that the results were generalizable to a different data set. The test set consisted of 20% of the total data, corresponding to 265 time series observations between April 2021 and April 2022. On the test set, the model had an adjusted $R^2$ value of 0.65, indicating very little model quality was lost and that the final model has significant generality.

# Conclusion

We have seen how influential NFTs are becoming across various different domains: art, entertainment, sports, academia, and more. In this paper, we have highlighted various methods of appraising an NFT for a specific collection. We showed how to use various different types of data sources to inform our models: public markets, search trends, and NFT metadata. We believe that we have aggregated the seminal works published thus far around NFT appraisals in the most sensible way to build a new SOTA NFT appraisal tool. Now that we have determined the key predictors that influence an NFT's price, these predictors can be fed into more advanced models (such as a multivariate RNN) to improve the quality of the predictions.

In this paper, we focused on the Bored Ape Yacht Club collection, which is a limitation of our research as we would like to explore the generalizability of this work to other projects. We believe that the work to be done with Twitter data needs to be explored more deeply given the relevance that Twitter places for NFT discoverability. We believe much more work can be done with language models and NFT appraisals that were out of scope for this project.

# Appendix

## Our codebase

Our development has been completed using Google Colab to share code among the team. The following links to a drive folder containing the code referenced within this report:
[Google Drive Code Repository](#)

## OpenSea API Data Dictionary

Upon calling the OpenSea API via [https://api.opensea.io/api/v1/assets](https://api.opensea.io/api/v1/assets), we are able to pull the following information about a given NFT:

['Project Name' 'one_day_volume' 'one_day_change' 'one_day_sales' 'one_day_average_price' 'seven_day_volume' 'seven_day_change' 'seven_day_sales' 'seven_day_average_price' 'thirty_day_volume' 'thirty_day_change' 'thirty_day_sales' 'thirty_day_average_price' 'total_volume' 'total_sales' 'total_supply' 'count' 'num_owners' 'average_price' 'num_reports' 'market_cap' 'floor_price']

The OpenSea API does not currently enable users to index historical data (you are only able to take a snapshot of the given day, the past seven days, and the past thirty days). In our explorations, we intend to use temporal data as a tool for future prediction which made it challenging to use the OpenSea API.

## Covalent API Data Dictionary

Upon calling [https://api.covalenthq.com/v1/1/nft_market/collection/](https://api.covalenthq.com/v1/1/nft_market/collection/), for a given collection, we are able to pull the following information about a given NFT.

{'average_volume_quote_day': 207.78058, 'average_volume_wei_day': '82503260869565200', 'chain_id': 1, 'collection_address': '0x2abb22d74dbc2b0f3c9bac9f173ef35ddb2c0809', 'collection_name': 'Geisha Tea House', 'collection_ticker_symbol': 'GTH', 'floor_price_quote_7d': 246.5085, 'floor_price_wei_7d': '97880923277777800', 'gas_quote_rate_day': 2518.453, 'opening_date': '2022-03-09', 'quote_currency': 'USD', 'unique_token_ids_sold_count_day': 46, 'volume_quote_day': 9557.906, 'volume_wei_day': '3795150000000000000'}.

The challenges we faced with the Covalent API was the way they have formatted their data. In the data dictionary, the floor price (the metric we are using for our label), is captured as an average of the floor price over the past seven days. The rest of the metrics (price of gas, average volume, etc.,) are captured as a daily input. This made it challenging to use daily metrics as a mechanism to predict a seven day average. However, at least with the Covalent API, we are able to turn all of the daily data into weekly averages by summing over a week's worth of data. The Covalent API requires more data cleaning but makes it amenable to work with for our objectives.

## Other NFT APIs explored

In addition to the OpenSea and Covalent API, we explored the Etherscan, Coingecko and Moralis APIs. We were constrained by the Coingecko API due to the lack of functionality they had around NFTs. Coingecko is a much better API to use for ERC-20 digital assets (typical cryptocurrencies and not ERC-721 NFTs). The Moralis API would force us to build within the Moralis framework making it hard for us to export code for the purposes of this assignment. Additionally, very early on, both Moralis charges money for its features after limited use. Etherscan has very poor documentation related to indexing NFT data making it hard for us to use.

## NFT Project Contract Addresses

Geisha Tea House: 0x2ABb22d74Dbc2B0F3C9BAC9f173ef35DdB2C0809'
Bored Ape Yacht Club: ,'0xBC4CA0EdA7647A8aB7C2061c2E118A18a936f13D','
Cryptopunks: 0xb47e3cd837dDF8e4c57F05d70Ab865de6e193BBB',
Doodles: '0x8a90CAb2b38dba80c64b7734e58Ee1dB38B8992e'
Azuki: ,'0xED5AF388653567Af2F388E6224dC7C4b3241C544',
Deadfellaz: '0x2acAb3DEa77832C09420663b0E1cB386031bA17B',
Gutter Cat Gang: '0xEdB61f74B0d09B2558F1eeb79B247c1F363Ae452'
Sup Ducks: ,'0x3Fe1a4c1481c8351E91B64D5c398b159dE07cbc5',
Cyber Kongs: '0x57a204AA1042f6E66DD7730813f4024114d74f37',
Creature World: '0xc92cedDfb8dd984A89fb494c376f9A48b999aAFc',
Cool Cats: '0x1A92f7381B9F03921564a437210bB9396471050C',
World of Women: '0xe785E82358879F061BC3dcAC6f0444462D4b5330',
Alien Frens: '0xd23d2D4aA76df5C4A19e1c9b6A83EA83f8c3db18',
Lazy Lions'0x8943C7bAC1914C9A7ABa750Bf2B6B09Fd21037E0'